\documentclass[fleqn]{article}
\usepackage{geometry}
\begin{document}

\date{}
\title{A Newman-Penrose Calculator for Instanton Metrics}
\author{T. Birkandan\thanks{
E-mail : \texttt{birkandant@itu.edu.tr}} \\
%EndAName
Istanbul Technical University, \\
Faculty of Science and Letters, \\
Dept. of Physics, Maslak 34469, Istanbul, Turkey}
\maketitle

\begin{abstract}
We present a Maple11+GRTensorII based symbolic calculator for instanton
metrics using Newman-Penrose formalism. Gravitational instantons are exact
solutions of Einstein's vacuum field equations with Euclidean signature. The
Newman-Penrose formalism, which supplies a toolbox for studying the exact
solutions of Einstein's field equations, was adopted to the instanton case
and our code translates it for the computational use.
\end{abstract}



\section{Introduction}

The interest in symbolic computational study of general relativity is
growing rapidly as the capacity of the the computer systems increase. The
computer is no longer an apparatus for the numerical relativist, but\ as the
symbolic manipulators get easier to use, the more researchers get into the
subject by using these systems. Comprehensive historical reviews can be
found in references \cite{cfa,his1,his2}. Another important point to
consider is the specialized packages which run under these platforms.
GRTensorII \cite{grtensor}, being a widespread package in general relativity
led significant progress in the area \cite%
{gror1,gror2,gror3,gror4,gror5,gror6}.

The Newman-Penrose (NP) formalism supplies a toolbox for investigating the
exact solutions of the Einstein's field equations \cite{np621,np622}.
Goldblatt has developed NP formalism for gravitational instantons based on
the $SU(2)\times SU(2)$ spin structure of positive definite metrics \cite%
{g941,g942}. In the gravitational instanton case, the gravitational field
decomposes into its self-dual and anti-self-dual parts and this
decomposition is natural in the spinor approach which necessitates two
independent spin frames for the spinor structure of 4-dimensional Rieman
manifolds with Euclidean signature \cite{an99}.

Gravitational instantons are exact solutions of Einstein's vacuum field
equations with Euclidean signature. They are analogous to the Yang-Mills
instantons, the finite action solutions of the classical Yang-Mills
equations. They admit hyper-K\"{a}hler structure \cite{an99,gr88}. For a
detailed overview one can see Eguchi et al's review and the articles cited
in this paper \cite{egh80}.

Aliev and Nutku applied differential forms to the NP formalism for
gravitational instantons which made the formalism more suitable for the
symbolic computation \cite{an99}.

\section{The program}

Our program, NPInstanton is designed for calculating physical and
mathematical quantities for instanton metrics using Newman-Penrose
formalism. It is coded under Maple 11 and GRTensorII package. The quantities
which can be calculated are:

- massless scalar equation,

- massless Dirac equation,

- source-free Maxwell equations,

- covariant and contravariant Dirac $\gamma $ matrices,

- coframe $l$($=l_{\mu }dx^{\mu }$) and $m$($=m_{\mu }dx^{\mu }$),

- Ricci rotation coefficients,

- Weyl scalars,

- trace-free Ricci scalars,

- spinor equivalent of the connection 1-forms,

- basis 2-forms,

- curvature 2-forms,

- integrands of the Euler number and the Hirzebruch signature curvature part
integrals,

- Petrov class of the space.

As one can see, some of the objects could be calculated by standard means
and without using a signature-dependent package. But for the sake of
completeness, we added these features to the program. By these, the program
becomes a complete symbolic calculator for an instanton metric. The NP
calculator of GRTensorII is not designed for the Euclidean signature and
could give unexpected results. Therefore, a complete NP based calculator,
combining the power of GRTensorII and NP formalism for these special metrics
is useful.

The program uses,

1. GRTensorII Package: Several objects (Ricci scalar, covariant Weyl tensor,
etc.) are calculated by this package in our program. No specific knowledge
of this package is needed for using NPInstanton. Our program creates the
metric file needed for the calculation itself using the NP legs given in the
input file and writes it to the metric directory of GRTensorII as
"npinstanton.mpl". This metric file then can be used in GRTensorII
independently.

2. DifferentialGeometry Package of Maple 11: The definition of the wedge
product is supplied by this package in our program. The wedge product
definition is also supplied by "difforms" package in Maple but it has
conflicts with linear algebraic quantities. This package is excluded in a
limited version of the program for the users of older Maple versions.

3. linalg package of Maple:\ This rather old internal package is used for an
eigenvalue calculation in Petrov classification section.

The program is set for a computer having a 512 Mb of RAM (Average value for
today's personal computers). If the system has less memory, the user must
change the line
\begin{verbatim}
kernelopts(gcfreq=10^7):
\end{verbatim}

of the \textbf{npinstanton.mws} file to a lower value (10\symbol{94}6 is the
standard value of Maple). For a computer having larger than 1 Gb of memory,
the user may change the gcfreq value as 10\symbol{94}8. "gc" is the
abbreviation of "garbage collection" and it is the Maple's internal routine
which cleans the memory after an amount of memory is allocated. For a
computer having a large amount of memory, one can increase the frequency of
this process. The larger the gc frequency value results in more memory to be
wasted but for a system having a large amount of memory it increases the
performance for some calculations.

The programming style is procedure-based. Each command is a Maple procedure
which can call other calculation procedures and inherit their outputs. For
example, when the \textbf{dirac() }command is given for the calculation of
massless Dirac equation, the program calls \textbf{dirac() }procedure. The
calculation of Dirac equations needs the information on $\gamma $ matrices.
Therefore, \textbf{dirac()} procedure calls \textbf{gammamatricespless()}
procedure for this calculation and its output is sent to \textbf{dirac()}
procedure. The suffix "\textbf{-pless}" (printless) means that this
procedure is a "secondary" one and it does not print its output to the
screen. When \textbf{dirac()} procedure finishes execution, it removes the
information that was inherited from \textbf{gammamatricespless()} procedure
and sends the user only its permitted output as \textbf{dirac} vector.

The "secondary" procedures do the massive works of the main procedures and
the main procedure generally contains the key directives. Therefore, this
usage makes the program easier to track. In addition to this, if a user
wants to add another command to the program, a procedure of that command can
be coded using these "secondary" procedures, or the user can add his/her own
"secondary" procedures, as well.

Throughout the program, only the output variables of the procedures are the
global ones and only these variables can be reached by the user at the end
of the execution of the command. Therefore, the user should be aware of the
output names of the procedures that are being used. This global variables
are shown at the end of the execution of each command. For example, the
output variable of \textbf{scalaroperator()} command will be shown as
\textbf{scalarop} and it can be called in the session when needed.

The outputs of the auxiliary calculations are cleared from the memory by
setting them as local variables, as these auxiliary calculations may occupy
too much memory. For example, $\gamma $ matrices information is not
necessary if the user needs to know only the Dirac equation. Therefore,
\textbf{dirac()} procedure clears the information about $\gamma $ matrices
after finishing the execution. If the user needs to know the $\gamma $
matrices, \textbf{gammamatrices()} procedure is run by the user. As another
example, in \textbf{conn1form()} procedure, only output values are
connection 1-forms and the outputs from \textbf{coframepless()} and \textbf{%
riccirotcoeffpless()} procedures, as well as internal calculations are
dismissed to gain memory for further calculations.

A few commands for simplification are added to the code which can be
evaluated easily by an average personal computer (i.e. having a 512 Mb of
memory), but it is always more convenient for the user to choose the right
simplification technique for the problem after the calculation. The output
must be regarded as a "raw material" for a simplification routine that is to
be chosen by the user. The user having a computer with insufficient memory
can extract the simplification routines from the program, simply by
modifying it in an editor but it is not recommended as it may lead to
miscalculation of some properties such as Petrov type or (anti-)self-duality.

One can reach the program files using the web site given in the "Final
Remarks" section. Linux and Windows versions are available and for those who
have older Maple versions, a limited version of NPInstanton which does not
contain the \textbf{basis2form()}, \textbf{curv2form()} and \textbf{%
topologicalnumbers()} parts (parts that use DifferentialGeometry Package) is
also supplied. The details on how to run the program is given in the README
file which comes with the program files.

\subsection{The input file}

The input file contains the necessary definitions for the space and wave
equations. General definitions (constants, etc.) can also be given in this
file. The definitions should be given using Maple's syntax. The input file
of our program plays the role of the metric file of GRTensorII.

The user first sets the coordinate names in the first section as,
\begin{verbatim}
firstcoordinate:= x:
secondcoordinate:= theta:
\end{verbatim}

etc. as given in example input file. Then the components of the covariant NP
legs are entered. \textbf{l\_covar[1]} being the first component of the
covariant $l_{\mu }$ leg and \textbf{l\_bar\_covar[1]} being the complex
conjugate of \ \textbf{l\_covar[1]}, etc.. Maple does not do the complex
simplifications because they need assumptions that may cause wrong
calculations. Therefore, the most appropriate way to define the legs is to
give both by hand. The choice of NP legs is not unique for a metric. For the
Eguchi-Hanson NP legs we can have\cite{egh80}:

\begin{eqnarray}
l &=&\frac{1}{\sqrt{2}}\left[ \frac{1}{\sqrt{1-\frac{a^{4}}{r^{4}}}}dr-\frac{%
ir}{2}\sqrt{1-\frac{a^{4}}{r^{4}}}(d\xi +\cos \theta d\phi )\right] ,
\label{eh_l} \\
m &=&\frac{re^{-i\xi }}{2\sqrt{2}}\left( d\theta +i\sin \theta d\phi \right)
.  \label{eh_m}
\end{eqnarray}

Then the input forms are
\begin{verbatim}
l_covar[1]:=1/(sqrt(2)*A):
l_bar_covar[1]:=1/(sqrt(2)*A):
\end{verbatim}

etc. as given in example input file. The $A$ value being $\sqrt{1-\frac{a^{4}%
}{r^{4}}}$ was given in the general definitions section.

Spinor components can be given if the user will be calculating the Dirac
equation. They can be chosen as \cite{su04},%
\begin{equation}
\psi _{1}=e^{i(m+\frac{1}{2})\xi }\Psi _{1}(r,\theta ,\phi ),
\end{equation}%
\begin{equation}
\psi _{2}=e^{i(m+\frac{1}{2})\xi }\Psi _{2}(r,\theta ,\phi ),
\end{equation}%
\begin{equation}
\psi _{3}=e^{i(m-\frac{1}{2})\xi }\Psi _{3}(r,\theta ,\phi ),
\end{equation}%
\begin{equation}
\psi _{4}=e^{i(m-\frac{1}{2})\xi }\Psi _{4}(r,\theta ,\phi )
\end{equation}

and the input form is
\begin{verbatim}
spinorcomponent1:=exp(I*(m+(1/2))*xi)*psi1[r,theta,phi]:
\end{verbatim}

etc. as given in example input file.

The scalar function can be given for the calculation of the scalar
equation. This definition can be skipped if the user does not need
to calculate this object. It can be chosen as

\begin{equation}
\varphi =e^{im\xi }\Phi (r,\theta ,\phi )
\end{equation}

for the Eguchi-Hanson space, $\varphi $ being the scalar function and it is
set as
\begin{verbatim}
scalarfunction:=exp(I*m*xi)*Phi(r,theta,phi):
\end{verbatim}

in the input file.

\subsection{Command definitions}

The list of commands can be given as,
\begin{verbatim}
scalaroperator()
dirac()
maxwell()
gammamatrices()
coframe()
spinrotcoeff()
weylscalar()
tfricciscalar()
conn1form()
basis2form()
curv2form()
topologicalnumbers()
\end{verbatim}

and the definitions of these commands are the following,

\begin{itemize}
\item scalaroperator():
\end{itemize}

This command calculates the massless scalar equation, finding the scalar
operator. The scalar operator is given as%
\begin{equation}
H\varphi \equiv \frac{1}{\sqrt{|g|}}\partial _{\nu }\sqrt{|g|}g^{\mu \nu
}\partial _{\mu }\varphi .
\end{equation}%
Here, $g$ is the determinant of the metric.

The procedure takes the scalar function (name: \textbf{scalarfunction}) from
the input file.

The output to be used thereafter:
\begin{verbatim}
> scalarop;
\end{verbatim}

For the massive case, one can equate this object to $M^{2}\varphi ^{2}$, $M$
being the mass of the scalar particle. As an additional property, this
procedure is not dependent on the metric signature and it does not use the
NP objects so it can be used for any space in four dimensions by extracting
it from the program.

\begin{itemize}
\item dirac():
\end{itemize}

This command calculates the massless Dirac equations,%
\begin{equation}
\gamma ^{\mu }\nabla _{\mu }\psi =0
\end{equation}%
where%
\begin{equation}
\nabla _{\mu }=\partial _{\mu }-\Gamma _{\mu }
\end{equation}%
and $\Gamma _{\mu }$ are the spin connections. The procedure takes the
spinor vector components from the input file under these names: \textbf{%
spinorcomponent1} ($\psi _{1}$), \textbf{spinorcomponent2} ($\psi _{2}$),
\textbf{spinorcomponent3} ($\psi _{3}$) and \textbf{spinorcomponent4} ($\psi
_{4}$). The output to be used thereafter is the components of the "\textbf{%
dirac}" vector as (i=1, 2, 3, 4):
\begin{verbatim}
> dirac[i];
\end{verbatim}

For the massive case, one can equate these objects to $\frac{M}{i}\psi $
vector, $M$ being the spinor mass and $i\equiv \sqrt{-1}$.

\begin{itemize}
\item maxwell():
\end{itemize}

This command calculates the source-free Maxwell equations using the
equations 102-109 of \cite{an99}. $F_{ij}$'s in the output are the usual
Maxwell field matrix components. The output is set to be "\textbf{maxwell}"
vector whose components are the source-free Maxwell equations as (i=1, 2, 3,
4):
\begin{verbatim}
> maxwell[i];
\end{verbatim}

\begin{itemize}
\item gammamatrices():
\end{itemize}

This command finds the covariant and contravariant Dirac $\gamma $ matrices
and tests them with anticommutation relations

\begin{equation}
\{\gamma ^{\mu },\gamma ^{\nu }\}=2g^{\mu \nu }.
\end{equation}

The output is "\textbf{gamma\_contr}" (contravariant $\gamma $ matrices:\ $%
\gamma ^{\mu }$) and "\textbf{gamma\_covar}" (covariant $\gamma $ matrices:\
$\gamma _{\mu }$) vectors as (i=1, 2, 3, 4):
\begin{verbatim}
> gamma_contr[i];
> gamma_covar[i];
\end{verbatim}

The calculation of the Dirac $\gamma $ matrices could be put into
the \textbf{dirac()} procedure but for further calculations
involving higher dimensions one may need only these matrices. To
find the $\gamma $ matrix for the extra dimension, one can use these
results and look for the $\gamma $ matrix of the extra dimension
using the anticommutation relations as in reference \cite{bh07}.
Therefore the procedure is a separate one.

\begin{itemize}
\item coframe():
\end{itemize}

This command calculates the coframe $l\equiv $\textbf{L} ($=l_{\mu }dx^{\mu
} $) and $m\equiv $\textbf{M} ($=m_{\mu }dx^{\mu }$). The output variables
are the following, "\textbf{\_bar}" denoting the complex conjugate:
\begin{verbatim}
> L;
> M;
> L_bar;
> M_bar;
\end{verbatim}

\begin{itemize}
\item riccirotcoeff():
\end{itemize}

This procedure calculates the Ricci rotation coefficients using equation 20
of \cite{an99}. The output can be used later by calling:
\begin{verbatim}
> npkappa; > nptau; > npsigma;
> nprho; > nppi; > npnu;
> npmu; > nplambda; > npgamma;
> npepsilon; > npalpha; > npbeta;
\end{verbatim}

in the meaning of the Ricci rotation coefficients $\kappa ,$ $\tau ,$ $%
\sigma ,$ $\rho ,$ $\pi ,$ $\nu ,$ $\mu ,$ $\lambda ,$ $\gamma ,$ $%
\varepsilon ,$ $\alpha ,$ $\beta $ respectively.

\begin{itemize}
\item weylscalar():
\end{itemize}

This command calculates the Weyl scalars using equation 68 of \cite{an99}.
The output variables are:
\begin{verbatim}
> weylscalar0; > weylscalar1; > weylscalar2, > weylscalar3;
> weylscalar4; > weylscalartilde0; >weylscalartilde1;
> weylscalartilde2; > weylscalartilde3; > weylscalartilde4;
\end{verbatim}

for $\Psi _{0}$, $\Psi _{1}$, $\Psi _{2}$, $\Psi _{3}$, $\Psi _{4}$, $\tilde{%
\Psi}_{0}$, $\tilde{\Psi}_{1}$, $\tilde{\Psi}_{2}$, $\tilde{\Psi}_{3}$, $%
\tilde{\Psi}_{4}$ respectively. "tilde" means the variable has a tilde in
the meaning of a different spin frame.

Another property of this procedure is that it finds out the Petrov class of
the space according to equations 120-122 of \cite{an99}. Petrov
classification of Euclidean spaces were first studied by Hacyan \cite{h79}
and then by Karlhede \cite{k86}.

\begin{itemize}
\item tfricciscalar():
\end{itemize}

This command calculates the trace-free Ricci scalars using equation 69 of
\cite{an99} and sets them to variables for a later use:
\begin{verbatim}
> tfricciscalar00; > tfricciscalar01; > tfricciscalar02;
> tfricciscalar10; > tfricciscalar11; > tfricciscalar12;
> tfricciscalar20; > tfricciscalar21; > tfricciscalar22;
> scalofcurv;
\end{verbatim}

for $\Phi _{00}$, $\Phi _{01}$, $\Phi _{02}$, $\Phi _{10}$, $\Phi _{11}$, $%
\Phi _{12}$, $\Phi _{20}$, $\Phi _{21}$, $\Phi _{22}$, scalar of curvature $%
\Lambda $ respectively.

\begin{itemize}
\item conn1form():
\end{itemize}

This procedure calculates the spinor equivalent of the connection 1-forms
given by equations 36-37 of \cite{an99}. The output can be reached by calling
\begin{verbatim}
> GAMMA00; > GAMMA01; > GAMMA10; > GAMMA11;
> GAMMAtilde0pr0pr; > GAMMAtilde0pr1pr;
> GAMMAtilde1pr0pr; > GAMMAtilde1pr1pr;
\end{verbatim}

for $\Gamma _{0}^{0}$, $\Gamma _{0}^{1}$, $\Gamma _{1}^{0}$, $\Gamma
_{1}^{1} $, $\tilde{\Gamma}_{0^{\prime }}^{0^{\prime }}$, $\tilde{\Gamma}%
_{0^{\prime }}^{1^{\prime }}$, $\tilde{\Gamma}_{1^{\prime }}^{0^{\prime }}$,
$\tilde{\Gamma}_{1^{\prime }}^{1^{\prime }}$ respectively where "tilde"
means the variable has a tilde and "\textbf{pr}" means "prime" in the
meaning of a different spin frame.

Another property of the procedure is that, it finds out the gauge by
checking the necessary and sufficient conditions for (anti-)self-duality
namely, $\Gamma _{ab}\equiv 0$ implies self duality and $\tilde{\Gamma}%
_{x^{\prime }y^{\prime }}\equiv 0$ implies anti-self-duality.

\begin{itemize}
\item basis2form():
\end{itemize}

This command finds the basis 2-forms using the definitions in equation 49 of
\cite{an99}. The output can be reached by
\begin{verbatim}
> L00; > L01; > L10;
> Ltilde0pr0pr; > Ltilde0pr1pr; > Ltilde1pr0pr;
\end{verbatim}

for $L_{0}^{0}$, $L_{0}^{1}$, $L_{1}^{0}$, $\tilde{L}_{0^{\prime
}}^{0^{\prime }}$, $\tilde{L}_{0^{\prime }}^{1^{\prime }}$, $\tilde{L}%
_{1^{\prime }}^{0^{\prime }}$ respectively. Here, "tilde" means the variable
has a tilde and "\textbf{pr}" means "prime" in the meaning of a different
spin frame.

\begin{itemize}
\item curv2form():
\end{itemize}

This command calculates the curvature 2-forms using equations 90-91 of \cite%
{an99} and sets them to these variables:
\begin{verbatim}
> Theta00; > Theta01; > Theta10;
> Thetatilde0pr0pr; > Thetatilde0pr1pr; > Thetatilde1pr0pr;
\end{verbatim}

for $\Theta _{0}^{0}$, $\Theta _{0}^{1}$, $\Theta _{1}^{0}$, $\tilde{\Theta}%
_{0^{\prime }}^{0^{\prime }}$, $\tilde{\Theta}_{0^{\prime }}^{1^{\prime }}$,
$\tilde{\Theta}_{1^{\prime }}^{0^{\prime }}$ respectively where "tilde"
means the variable has a tilde and "\textbf{pr}" means "prime" in the
meaning of a different spin frame.

\begin{itemize}
\item topologicalnumbers():
\end{itemize}

This command calculates the integrands of the Euler number and the
Hirzebruch signature curvature part integrals using the relations 115-117 of
\cite{an99} namely,

\begin{eqnarray}
\chi &=&\frac{1}{4\pi ^{2}}\int_{\mathcal{M}}\left[ \,|\,\Psi
_{0}\,|^{2}+4\,|\,\Psi _{1}\,|^{2}+3\,\Psi _{2}^{2}+|\,\tilde{\Psi}%
_{0}\,|^{2}\right.  \nonumber \\[0.12in]
&&\left. +\,4\,|\,\tilde{\Psi}_{1}\,|^{2}+3\,\tilde{\Psi}_{2}^{2}-2\,(\,|\,%
\Phi _{00}\,|^{2}+|\,\Phi _{02}\,|^{2}\,)\right.  \nonumber \\[0.12in]
&&\left. -\,4\,(\,|\,\Phi _{01}\,|^{2}+|\,\Phi _{11}\,|^{2}+|\,\Phi
_{12}\,|^{2}-3\Lambda ^{2}\,)\,\right] l\wedge \bar{l}\wedge m\wedge \bar{m}
\end{eqnarray}%
\begin{eqnarray}
\tau &=&-\frac{1}{6\pi ^{2}}\int_{\mathcal{M}}\left[ \,|\,\Psi
_{0}\,|^{2}+4\,|\,\Psi _{1}\,|^{2}+3\,\Psi _{2}^{2}-|\,\tilde{\Psi}%
_{0}\,|^{2}\right.  \nonumber \\[0.12in]
&&\left. -\,4\,|\,\tilde{\Psi}_{1}\,|^{2}-3\,\tilde{\Psi}_{2}^{2}\,\right]
l\wedge \bar{l}\wedge m\wedge \bar{m}\,-\eta _{s}({\partial \mathcal{M}})
\end{eqnarray}

$\eta _{s}({\partial \mathcal{M}})$ being the eta-invariant and this value
will not be taken into consideration in the program. The output can be
reached by calling
\begin{verbatim}
> eulernumber_integrand;
> hirzebruch_signature_integrand;
\end{verbatim}

These numbers have a special importance for the Atiyah-Patodi-Singer index
theorem of operators on manifolds with boundary \cite{aps751}\cite{aps752}%
\cite{hrs80}\cite{h83}.

\section{Examples}

In this section, we will apply our program to two instanton metrics and
calculate some objects using the special commands. Lengthy output values are
not written to avoid distracting the reader's attention.

\subsection{Example 1: Calculations for the Eguchi-Hanson metric}

Eguchi-Hanson instanton \cite{egh80} is the most similar to the Yang-Mills
instanton of Belavin et al. \cite{bpst75} and the metric is given as

\begin{equation}
ds^{2}=\frac{1}{1-\frac{a^{4}}{r^{4}}}dr^{2}+r^{2}(\sigma _{x}^{2}+\sigma
_{y}^{2})+r^{2}(1-\frac{a^{4}}{r^{4}})\sigma _{z}^{2}.
\end{equation}

Here,%
\begin{eqnarray}
\sigma _{x} &=&\frac{1}{2}(-\cos \xi d\theta -\sin \theta \sin \xi d\phi ),
\nonumber \\
\sigma _{y} &=&\frac{1}{2}(\sin \xi d\theta -\sin \theta \cos \xi d\phi ), \\
\sigma _{z} &=&\frac{1}{2}(-d\xi -\cos \theta d\phi ).  \nonumber
\end{eqnarray}

and the dyad was given in eq. \ref{eh_l} and eq. \ref{eh_m}. We run our
program in Maple and type the name of the input file: \textbf{eguchihanson}

Now, let us calculate the connection 1-forms:
\begin{verbatim}
npinstanton> conn1form();
\end{verbatim}

The program calculates and shows the calculated values. Then the gauge is
found by the program as:
\begin{verbatim}
SELF DUAL GAUGE because all connection 1-forms without
tilde are zero
\end{verbatim}

before prompting for another calculation, the procedure shows the output
variables which can be used for a later calculation. for example, let us
call $\Gamma _{0}^{1}$ of equation 36 of \cite{an99}:
\begin{verbatim}
npinstanton> GAMMA01;
\end{verbatim}

To find the Weyl scalars and the Petrov class, one can run,
\begin{verbatim}
npinstanton> weylscalar();
\end{verbatim}

After the Weyl scalars are shown, the Petrov type is found to be:
\begin{verbatim}
Petrov-type D according to anti-self-dual part
\end{verbatim}

All these results agree with the literature \cite{k86}.

We can find the scalar operator by using
\begin{verbatim}
npinstanton> scalaroperator()
\end{verbatim}

command. The scalar equation can be solved in terms of hypergeometric
equations in four dimensional case by the command,
\begin{verbatim}
npinstanton> pdsolve(scalarop,INTEGRATE);
\end{verbatim}

We can easily find the solution for the scalar equation with one extra
dimension added trivially if this dimension is a Killing direction (i.e. $%
ds^{2}=-dt^{2}+ds_{4}^{2}$, $ds_{4}^{2}$ being the original instanton metric
similar to the case in reference \cite{bh07}). As an example, we just add $%
-k_{t}^{2}\Phi (r,\theta )$ to the equation after dividing by the
exponential part which comes from the solution of the Killing directions:
\begin{verbatim}
npinstanton> pdsolve(scalarop/(exp((m*xi+n*phi)*I))
             -kt*kt*Phi(r,theta),INTEGRATE);
\end{verbatim}

Here, $k_{t}$ is the eigenvalue corresponding to the extra dimension. We see
that the addition of the extra dimension results in a more singular
equation, confluent Heun equation as the solution of the radial part. The
occurrence of Heun equations in higher dimensional solutions is known in the
literature (see \cite{bh07} and references therein) and our example is a new
one.

\subsection{Example 2: Calculations for the Nutku helicoid metric}

The NP legs of the Nutku helicoid metric can be chosen as \cite{ahkn99},

\begin{equation}
l^{\mu }=\frac{a\sqrt{\sinh 2x}}{2}(1,i,0,0),
\end{equation}
\begin{equation}
m^{\mu }=\frac{1}{\sqrt{\sinh 2x}}(0,0,\cosh (x-i\theta ),i\sinh (x-i\theta
)).
\end{equation}

This is an example of a multi-center metric. This metric reduces to
the flat metric if we take $a=0$. Since this solution has curvature
singularities, it has not been studied extensively aside from four
articles \cite{su04}\cite{bh07}\cite{bh07II}\cite{v05}.

Let us calculate the basis 2-forms:
\begin{verbatim}
npinstanton> basis2form();
\end{verbatim}

After the calculation, to call $L_{0}^{0}$ of equation 49 of \cite{an99},
\begin{verbatim}
npinstanton> L00;
\end{verbatim}

and the output will be

-$\frac{1}{2}Ia^{2}\cosh (x)\sinh (x)dx1\wedge dx2$-$\frac{1}{2}Idx3\wedge
dx4$

$dx$'s were defined previously, at the very beginning of the session in
terms of the real coordinates. We can calculate the Dirac equation by,
\begin{verbatim}
npinstanton> dirac();
\end{verbatim}

and
\begin{verbatim}
npinstanton> dirac[3];
\end{verbatim}

calls the third component of the Dirac equation vector which can be
simplified and used for calculations.

We can calculate the connection 1-forms as,
\begin{verbatim}
npinstanton> conn1form();
\end{verbatim}

For this definition the gauge turns out to be:
\begin{verbatim}
ANTI-SELF DUAL GAUGE because all connection 1-forms with
tilde are zero
\end{verbatim}

and the connection 1-forms are shown.

For the Weyl scalars and the Petrov class, one can run,
\begin{verbatim}
npinstanton> weylscalar();
\end{verbatim}

After the Weyl scalars are shown on the screen, the Petrov type is found to
be:
\begin{verbatim}
Petrov-type I according to self-dual part
\end{verbatim}

The whole results agree with the literature \cite{ahkn99}.

Now let us take a metric of the form $ds^{2}=-dt^{2}+ds_{4}^{2}$, $%
ds_{4}^{2} $ being the original Nutku helicoid metric. In this case, we need
one more $\gamma $ matrix for the extra dimension ($\gamma ^{t}$). Firstly,
let us use our command to find four dimensional $\gamma $ matrices:
\begin{verbatim}
npinstanton> gammamatrices();
\end{verbatim}

It is clear that the extra $\gamma $ matrix will be diagonal as the four
dimensional ones are non-diagonal. Then we can write these commands to form
equations using $\{\gamma ^{\mu },\gamma ^{\nu }\}=2g^{\mu \nu }:$
\begin{verbatim}
extragamma:=Matrix([[ x1, x2,  0,  0 ],[ x3, x4,  0,  0 ],
                    [ 0,  0,  x5, x6 ],[ 0,  0,  x7, x8 ]]):
\end{verbatim}

We have eight unknowns, so we will have eight equations:
\begin{verbatim}
eqn1:=(gamma_contr[1].extragamma+extragamma.gamma_contr[1])[1,3];
eqn2:=(gamma_contr[1].extragamma+extragamma.gamma_contr[1])[1,4];
eqn3:=(gamma_contr[1].extragamma+extragamma.gamma_contr[1])[2,3];
eqn4:=(gamma_contr[1].extragamma+extragamma.gamma_contr[1])[2,4];
eqn5:=(gamma_contr[2].extragamma+extragamma.gamma_contr[2])[1,3];
eqn6:=(gamma_contr[2].extragamma+extragamma.gamma_contr[2])[1,4];
eqn7:=(gamma_contr[2].extragamma+extragamma.gamma_contr[2])[2,3];
eqn8:=(gamma_contr[2].extragamma+extragamma.gamma_contr[2])[2,4];
\end{verbatim}

and we solve them,
\begin{verbatim}
sol:=solve({eqn1=0,eqn2=0,eqn3=0,eqn4=0,
            eqn5=0,eqn6=0,eqn7=0,eqn8=0}
          ,{x1,x2,x3,x4,x5,x6,x7,x8});
\end{verbatim}

Then, we have the form of the unknown matrix:
\begin{verbatim}
preresult:=subs(sol,extragamma);
\end{verbatim}

To satisfy $\{\gamma ^{t},\gamma ^{t}\}=-2$, we can set $x5=i,x8=-i$:
\begin{verbatim}
extramatrix:=subs(x5=I,x8=-I,preresult);
\end{verbatim}

Then we can use this matrix with the name "\textbf{extramatrix}".

\section{Final remarks}

We introduced a Maple11+GRTensorII based symbolic calculator consisting of
procedures for instanton metrics using a Euclidean Newman-Penrose formalism.
A limited version which can run in older versions of Maple is also available.

As the program consists of procedures for different calculations,
development of the program according to the user's needs is easy by adding
procedures.

The program and sample files can be downloaded from the address:
\begin{verbatim}
https://github.com/tbirkandan/NPInstanton
\end{verbatim}

\section*{Acknowledgement}

The author would like to express his sincere to thanks to Prof. Mahmut Horta%
\c{c}su for help and support and Profs. Ay\c{s}e H. Bilge and Ne\c{s}e \"{O}%
zdemir for scientific assistance and Fuat \.{I}. Vardarl\i\ for technical
assistance throughout this work. This work is supported by TUBITAK, the
Scientific and Technological Council of Turkey.

\end{document}